\documentclass[onecolumn, twoside, 12pt]{IEEEtran}
\usepackage{amsmath, amssymb,graphicx,psfrag,cite}

\newtheorem{theorem}{Theorem}
\newtheorem{definition}{Definition}
\newtheorem{corollary}{Corollary}
\newtheorem{lemma}{Lemma}

\begin{document}

\setcounter{page}{0}


\title{Construction of Near-Optimum Burst Erasure Correcting
Low-Density Parity-Check Codes}
\author{
    Enrico~Paolini,~\IEEEmembership{Member,~IEEE}, and
    Marco~Chiani,~\IEEEmembership{Senior~Member,~IEEE} \\
\thanks{This research was supported in part by ESA/ESOC.}
\thanks{The authors are with DEIS/WiLAB,
        University of Bologna,
        Via Venezia 52,
        47023 Cesena (FC), ITALY
        (e-mail: {\tt e.paolini@unibo.it}, {\tt marco.chiani@unibo.it}).}
}
\markboth
    {IEEE Transactions on Communications, to appear}
    {Construction of Near-Optimum Burst Erasure Correcting Low-Density Parity-Check Codes}
%
\maketitle

\thispagestyle{empty}

\setcounter{page}{1}

\begin{abstract}

In this paper, a simple, general-purpose and effective tool for
the design of low-density parity-check (LDPC) codes for iterative
correction of bursts of erasures is presented. The design method
consists in starting from the parity-check matrix of an LDPC code
and developing an optimized parity-check matrix, with the same
performance on the memory-less erasure channel, and suitable also
for the iterative correction of single bursts of erasures. The
parity-check matrix optimization is performed by an algorithm
called pivot searching and swapping (PSS) algorithm, which
executes permutations of carefully chosen columns of the
parity-check matrix, after a local analysis of particular variable
nodes called stopping set pivots. This algorithm can be in
principle applied to any LDPC code. If the input parity-check
matrix is designed for achieving good performance on the
memory-less erasure channel, then the code obtained after the
application of the PSS algorithm provides good joint correction of
independent erasures and single erasure bursts. Numerical results
are provided in order to show the effectiveness of the PSS
algorithm when applied to different categories of LDPC codes.

\end{abstract}


\section{Introduction}\label{sec:intro}
Recently, the problem of designing low-complexity codes for
transmission on burst erasure channels, especially low-density
parity-check (LDPC) codes \cite{gallager63:low-density}, has
gained a certain interest (e.g.
\cite{ryan04:epr4,ryan06:simplified,Calzolari2007,paolini06:spaceops,wadayama06:ensemble,tai06:burst,tai06:burst-journal,divsalar06:burst,paolini06:improved,hosoya06:performance,hosoya06:modification,wadayama04:greedy,johnson04:bursty}).
This demand for burst erasure correcting codes can be explained by
the fact that such codes are quite interesting for several
applications, including magnetic storage, wireless communications
on correlated fading channels, and even space communications. For
instance, LDPC codes with good properties in terms of correction
of single bursts of erasures have been shown in \cite{ryan04:epr4}
and \cite{ryan06:simplified} to represent a promising solution,
respectively, for the error control system in magnetic recording
applications, and for communication on correlated Rayleigh fading
channels. Furthermore, it is currently under investigation the
possibility to implement LDPC-like codes at the upper layers of
the communication stack, for correcting bursts of packet erasures
in space and satellite communication links
\cite{Calzolari2007,paolini06:spaceops}, in order to heavily
reduce the use of automatic repeat request (ARQ) protocols.

It has been shown in \cite{fossorier05:universal} that practically
any $(n,k)$ linear block code can be used to correct any single
burst of $n-k$ or less erasures, thus achieving the optimal
correction capability of single bursts of erasures.
More specifically, it has been proved that, under very mild
assumptions, it is possible to obtain a (redundant) representation
of the parity-check matrix for the code (called parity-check
matrix in burst correction form) that permits to recover from any
pattern of $n-k$ or less contiguous erasures, by applying a
decoding algorithm whose computational complexity is quadratic in
the codeword length $n$.
The same code can be then used in a communication system for
erasure recovery both in scenarios with independent bit erasures
and in scenarios where the erasures occur in bursts. The decoding
is performed according to a two-step process: the received
sequence is first processed by the burst correcting algorithm
operating on the parity-check matrix in burst correction form, and
then by the decoding algorithm for independent erasures
correction, operating on a different parity-check matrix
representation.

This very general technique can be in principle applied also to
LDPC codes.
In this case, in the first step of the decoding process, the burst
erasure correcting decoder is applied, with quadratic complexity,
to the parity-check matrix in burst correction form; in the second
step of the decoding process, the iterative decoding
\cite{luby01:efficient} is performed on a low-density
representation of the parity-check matrix.

In this paper, the possibility to construct LDPC codes, capable to
perform recovery of both independent erasures and bursts of
erasures by exploiting \emph{only} the iterative decoder, is
investigated.
More specifically, the approach consists in starting from an LDPC
code parity-check matrix, usually designed in order to achieve
good performance on the (memory-less) binary erasure channel
(BEC\footnote{Throughout the paper the acronym BEC will be used to
denote the standard memory-less binary erasure channel.}), and
then properly modifying it in order to make it suitable also for
the \emph{iterative} correction of single bursts of erasures.

The performance of this single-step LDPC iterative decoding
process is in general suboptimal, in burst scenarios, with respect
to the two-step technique proposed in
\cite{fossorier05:universal}, because of the suboptimal iterative
burst correction. However, avoiding the quadratic complexity burst
correction step (which becomes an issue for long LDPC codes) it
only requires linear in $n$ decoding complexity
\cite{luby01:efficient}.
Furthermore, the proposed algorithm used for making the
parity-check matrix suitable to iterative correction of bursts of
erasures, besides being extremely simple, turns out to be also
extremely effective, generating finite length LDPC codes whose
erasure burst correction capability is very close to its maximum
possible value. Moreover, if the input parity-check matrix is
designed for achieving good performance on the (memory-less) BEC,
the optimized code can be used for transmission in both burst and
independent erasure scenarios.

The present paper is strictly related to a number of recent works,
i.e.
\cite{ryan04:epr4,ryan06:simplified,wadayama06:ensemble,divsalar06:burst,paolini06:improved}.
In \cite{ryan04:epr4}, a key parameter is proposed as a measure of
the robustness of an LDPC code to single bursts of erasures,
namely the \emph{maximum guaranteed resolvable erasure burst
length}, denoted by $L_{\max}$.
A general definition of the parameter $L_{\max}$, valid for any
code, can be given as follows.
\medskip
\begin{definition}
For a given code, a given parity-check matrix representation, and
a given decoding algorithm, the maximum guaranteed resolvable
erasure burst length ($L_{\max}$) is the maximum length of an
erasure burst which is correctable independently of its position
within the codeword.
\end{definition}
\medskip
As explicitly remarked in this definition, $L_{\max}$ is not
unique for a given code, heavily depending on both the decoding
algorithm and the parity-check matrix representation. For
instance, for a given parity-check matrix representation of an
LDPC code, the value of $L_{\max}$ is not the same with respect to
the standard iterative decoding algorithm or to the improved
decoding algorithm proposed in \cite{fekri04:decoding}. In the
sequel, the standard iterative decoder for LDPC codes will be
always considered.
In \cite{ryan04:epr4}, an algorithm for the efficient computation
of $L_{\max}$ for LDPC codes under iterative decoding is
developed.

An estimate of the optimal value of $L_{\max}$ for LDPC codes,
under standard iterative decoding, has been proposed in
\cite{ryan06:simplified}. Consider an LDPC code, and let $p^*$
denote the associated asymptotic decoding threshold
\cite{richardson01:design} on the BEC. Then, for sufficiently
large codeword length $n$, there exists some proper permutation of
the parity-check matrix columns such that $L_{\max}/n \simeq p^*$.
Then, $\lfloor p^* n \rfloor$ can be used as an estimate of the
maximum value of $L_{\max}$ that can be obtained for the
length-$n$ LDPC code, by permuting the parity-check matrix
columns.

The maximum guaranteed resolvable erasure burst length for an LDPC
code, under iterative decoding, has a strong dependence on the
\emph{stopping sets} present in the bipartite graph. The concept
of stopping set has been first introduced in \cite{di02:finite}.
By definition, a stopping set is any set of variable nodes such
that any check node connected to this set is connected to it at
least twice. In \cite{di02:finite}, it is also proved that the
union of stopping sets is a stopping set, so that it is possible
to define a maximal stopping set included in a subset of the
variable nodes, as the union of all the stopping sets included in
the subset. The residual erased variable nodes (after iterative
decoding) constitute the maximal stopping set included in the
original erasure pattern. Hence, a decoding failure takes place
whenever the erasure pattern due to the channel contains a
stopping set.

The relation between stopping sets and $L_{\max}$ has been
addressed in \cite{wadayama06:ensemble}. Let $\mathcal{G}$ be the
LDPC code bipartite graph, and let $\mathcal{V}=\{ V_0, V_1,
\dots, V_{n-1} \}$ be the set of the variable nodes. If
$\mathcal{S}=\{ V_{i_1},V_{i_2},\dots,V_{i_t} \}$, with
$i_1<i_2<\dots<i_t$, is a stopping set, then the span of
$\mathcal{S}$ is defined as $1+|i_t-i_1|$. Denoting by
$\mu(\mathcal{G})$ the minimum span of stopping sets (i.e. the
minimum among the spans of all the stopping sets of
$\mathcal{G}$), it follows that $L_{\max}=\mu(\mathcal{G})-1$. The
span of the stopping sets, a concept of no interest on the BEC,
heavily affects the code performance when the erasures occur in
bursts.

The concept of span of stopping sets is considered in
\cite{tai06:burst-journal}, where a lower bound is found for the
minimum span of stopping sets for any regular LDPC code. More
specifically, it is proved that the minimum span of stopping sets
satisfies $\mu(\mathcal{G}) \geq \delta$, where $\delta$ is the
minimum \emph{zero span} in the parity-check matrix. A zero-span
is defined as a sequence of consecutive zeroes in a parity-check
matrix row; in terms of $L_{\max}$ the bound is $L_{\max} \geq
\delta - 1$. In \cite{tai06:burst-journal}, a technique for
constructing regular LDPC codes with good $L_{\max}$ is also
presented. A class of protograph-based LDPC codes whose minimum
stopping set size increases linearly with the codeword length $n$,
and hence suitable for transmission over burst erasure channels,
has been presented in \cite{divsalar06:burst}.

The main contribution of this paper is the development of a greedy
algorithm, which is able to modify the parity-check matrix of an
LDPC code, designed for erasure correction on the BEC, in order to
make it suitable also for the iterative correction of single
erasure bursts. The present work extends and improves some results
from \cite{paolini06:improved}, where a former version of the
algorithm was presented.
The developed algorithm is called pivot searching and swapping
(PSS) algorithm, since it is based on the search and swap of the
\emph{pivots} of stopping sets. The novel concept of pivot of a
stopping set will be introduced in the next section.
According to the proposed approach, the parity-check matrix for
iterative burst correction is generated by only performing column
permutations on the input parity-check matrix.
Hence, the sparseness of the matrix is not altered by the
algorithm, and the iterative decoder applied to the new
parity-check matrix leads, on the BEC, to exactly the same
performance as the original one.
If the parity-check matrix received in input by the algorithm is
designed for achieving good performance on the BEC, then the code
obtained after the application of the PSS algorithm provides good
joint correction of independent erasures and single erasure
bursts, within a single-step decoding scheme, only exploiting the
iterative decoder.

A related algorithm, developed in \cite{wadayama04:greedy},
combines column permutations with column eliminations to improve
the code $L_{\max}$. On the other hand, as remarked above, only
column permutations are performed by the algorithm proposed in
this paper, based on the concept of stopping set pivot. A second
related algorithm, based on permutations of the parity check
matrix columns, has been developed in \cite{hosoya06:modification}
for obtaining improved LDPC codes for the correction of two or
more bursts of erasures. The metric adopted in
\cite{hosoya06:modification} to measure the code performance is
based on the concepts of average distance between elements and
minimum distance between elements, and is different from
$L_{\max}$. We also point out the LDPC code construction
technique, based on circulant matrices, proposed in
\cite{johnson04:bursty} and aimed at obtaining LDPC codes with a
good compromise between correction of random erasures and erasure
bursts. Differently from the approach in \cite{johnson04:bursty},
the algorithm developed in this paper can be applied to improve
the robustness to a single erasure burst of any LDPC code.

The paper is organized as follows. In Section \ref{sec:ss-pivots}
the concept of pivot of a stopping set is presented, some
properties of the pivots of stopping sets are proved, and an
efficient algorithm for finding some pivots of a given stopping
set is proposed. Section \ref{sec:optimization-algorithm} is
devoted to the detailed description of the PSS algorithm. In
Section \ref{sec:numerical-results} some numerical results are
presented, showing the improvement in terms of $L_{\max}$
achievable by applying the PSS algorithm to different types of
LDPC codes. Finally, Section \ref{sec:conclude} concludes the
paper.

\section{Pivots of Stopping Sets}\label{sec:ss-pivots}
In this section, the novel concept of pivots of a stopping set is
introduced. It is given proof that the minimum number of pivots
for any stopping set is two, and an efficient algorithm for
finding some pivots of a given stopping set is developed.

For any LDPC code, and for any stopping set of the LDPC code, we
define \emph{subgraph induced by the stopping set} the bipartite
graph composed of the variable nodes which are part of the
stopping set, the check nodes connected (necessarily at least
once) to these variable nodes, and the edges connecting such
variable nodes and such check nodes. The key concept of
\emph{pivot} of a stopping set is defined next.

\medskip
\begin{definition}\label{def:pivot}
Let $\mathcal{G}$ be the subgraph induced by a stopping set
$\mathcal{S}$ of an LDPC code. A variable node $V$ is called pivot
of the stopping set if the following property holds: if the value
of $V$ is known and the value of all the other variable nodes of
$\mathcal{G}$ is unknown, then the iterative decoder applied to
$\mathcal{G}$ is able to successfully recover from the erasure
pattern.
\end{definition}

\medskip
According to this definition, if the variable node $V$ is pivot
for a stopping set $\mathcal{S}$, then the set of variable nodes
$\mathcal{S} / \{ V \}$ is not a stopping set for the LDPC code
and contains no stopping sets.

As recalled in the previous section, the iterative decoder is not
able to recover from a starting erasure pattern caused by the
channel when this erasure pattern includes at least one stopping
set. In particular, the set of variable nodes which remain
uncorrected at the end of the decoding process is the maximal
stopping set included in the starting erasure pattern, i.e. the
union of all the stopping sets included in it. The residual graph
at the end of the decoding process is the subgraph induced by the
stopping set. What we point out with the above definition is that,
among the variable nodes in this residual graph, one should
distinguish between pivot and non-pivot variable nodes: if the
value of at least one of the pivots was known, then the decoding
would be successful. This is the basic idea exploited in the
optimization algorithm presented in the next section.

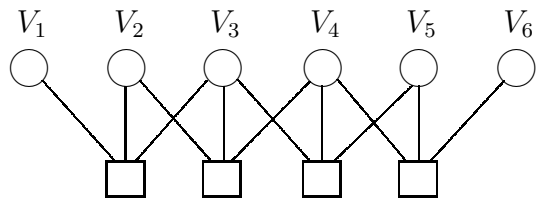
\begin{figure}[!h]
\centerline{
\unitlength .7mm 
\linethickness{0.4pt}
\ifx\plotpoint\undefined\newsavebox{\plotpoint}\fi 
\begin{picture}(101.15,34.75)(80,110)
\put(103.75,136.25){\circle{6.8}}
\put(85.25,136.25){\circle{6.8}}
\put(141,136.25){\circle{6.8}}
\put(122,136.25){\circle{6.8}}
\put(159.25,136.25){\circle{6.8}}
\put(177.75,136.25){\circle{6.8}}
\put(104.25,144.75){\makebox(0,0)[cc]{$V_2$}}
\put(85.75,144.75){\makebox(0,0)[cc]{$V_1$}}
\put(141.5,144.75){\makebox(0,0)[cc]{$V_4$}}
\put(122.5,144.75){\makebox(0,0)[cc]{$V_3$}}
\put(159.75,144.75){\makebox(0,0)[cc]{$V_5$}}
\put(178.25,144.75){\makebox(0,0)[cc]{$V_6$}}
\put(100,112){\framebox(7,6.5)[cc]{}}
\put(137.25,112){\framebox(7,6.5)[cc]{}}
\put(118.25,112){\framebox(7,6.5)[cc]{}}
\put(155.5,112){\framebox(7,6.5)[cc]{}}
\put(159.25,118.5){\line(0,1){14.5}}
\multiput(161.25,118.5)(.048141892,.052364865){296}{\line(0,1){.052364865}}
\multiput(123.5,118.5)(.048076923,.048878205){312}{\line(0,1){.048878205}}
\multiput(124.75,134.25)(.048076923,-.054195804){286}{\line(0,-1){.054195804}}
\multiput(143.75,134.25)(.048148148,-.057407407){270}{\line(0,-1){.057407407}}
\put(140.75,133){\line(0,-1){14.25}}
\multiput(158,133)(-.049831081,-.048141892){296}{\line(-1,0){.049831081}}
\multiput(106.25,134)(.048181818,-.054545455){275}{\line(0,-1){.054545455}}
\multiput(87.75,134)(.048042705,-.053380783){281}{\line(0,-1){.053380783}}
\put(103.5,132.75){\line(0,-1){13.75}}
\put(122.25,132.75){\line(0,-1){14}}
\multiput(105.5,118.75)(.048042705,.054270463){281}{\line(0,1){.054270463}}
\end{picture}} \caption{Example of
subgraph induced by a stopping set of size 6, with no pivots. }
\label{fig:SS-without-pivots}
\end{figure}

\begin{figure}[!h]
\centerline{
\unitlength .7mm 
\linethickness{0.4pt}
\ifx\plotpoint\undefined\newsavebox{\plotpoint}\fi 
\begin{picture}(64.4,34.75)(80,110)
\put(103.75,136.25){\circle{6.8}}
\put(85.25,136.25){\circle{6.8}}
\put(141,136.25){\circle{6.8}}
\put(122,136.25){\circle{6.8}}
\put(104.25,144.75){\makebox(0,0)[cc]{$V_2$}}
\put(85.75,144.75){\makebox(0,0)[cc]{$V_1$}}
\put(141.5,144.75){\makebox(0,0)[cc]{$V_4$}}
\put(122.5,144.75){\makebox(0,0)[cc]{$V_3$}}
\multiput(96.25,119)(.04782609,.12608696){115}{\line(0,1){.12608696}}
\multiput(119.75,133.5)(-.04816514,-.13302752){109}{\line(0,-1){.13302752}}
\multiput(124.25,133.5)(.04807692,-.11153846){130}{\line(0,-1){.11153846}}
\put(90.75,111.75){\framebox(7.5,7.25)[]{}}
\put(108.5,111.75){\framebox(7.5,7.25)[]{}}
\put(127,111.75){\framebox(7.5,7.25)[]{}}
\multiput(87,133)(.077797203,-.048076923){286}{\line(1,0){.077797203}}
\multiput(85,133)(.04801325,-.09271523){151}{\line(0,-1){.09271523}}
\multiput(88.5,135)(.118222892,-.048192771){332}{\line(1,0){.118222892}}
\multiput(105.25,133)(.04791667,-.11458333){120}{\line(0,-1){.11458333}}
\multiput(140.25,133)(-.04801325,-.09271523){151}{\line(0,-1){.09271523}}
\end{picture}} \caption{Example of
subgraph induced by a stopping set of size 4. The variable nodes
$V_1$ and $V_2$ are pivots for the stopping set, the variable
nodes $V_3$ and $V_4$ are not pivots.} \label{fig:SS-with-pivots}
\end{figure}
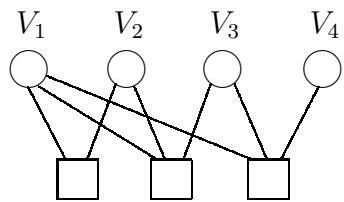

It is important to underline that not all the stopping sets have
pivots. For instance, the stopping set of size 6 with the induced
subgraph depicted in Fig. \ref{fig:SS-without-pivots} has no
pivots, while the stopping set of size 4 with the induced subgraph
depicted in Fig. \ref{fig:SS-with-pivots} has two pivots, $V_1$
and $V_2$. The concept of \emph{span of pivots}, defined next,
will be used in Section \ref{sec:optimization-algorithm}.

\medskip
\begin{definition}
Let $\mathcal{S}$ be a stopping set with pivots, and let $V_p$ and
$V_q$ be, respectively, the pivot of $\mathcal{S}$ with minimum
index and the pivot of $\mathcal{S}$ with maximum index. Then, the
span of the pivots of $\mathcal{S}$ is then defined as
$q-p+1$~\footnote{The fact that a stopping set with pivots has at
least two pivots will be proved in Theorem
\ref{theo:at-least-2-pivots}.}.
\end{definition}

\medskip
The next lemma points out an important property of the structure
of stopping sets characterized by the presence of pivots.

\medskip
\begin{lemma}\label{lemma:non-disjoint}
No stopping set $\mathcal{S}$ with pivots exists whose induced
subgraph is composed by disjoint graphs.
\end{lemma}
\begin{proof}
It is well known that the union of stopping sets is a stopping set
\cite{di02:finite}. Hence, the union of stopping sets with
unconnected induced subgraphs is a stopping set. Let $\mathcal{G}$
be the subgraph induced by a stopping set, and let $\mathcal{G} =
\mathcal{G}_1 \cup \mathcal{G}_2$, with $\mathcal{G}_1$ and
$\mathcal{G}_2$ unconnected. In such a condition, even if a
variable node $V_\alpha$ is pivot with respect to $\mathcal{G}_1$,
it cannot be pivot for the whole stopping set, because no variable
node in $\mathcal{G}_2$ can be corrected from the knowledge of the
value of $V_\alpha$ only. Analogously, even if a variable node
$V_\beta$ is pivot with respect to $\mathcal{G}_2$, it cannot be
pivot for the whole stopping set. Hence, the stopping set has no
pivots.
\end{proof}

\medskip
For a given stopping set of an LDPC code, the problem of finding
all the stopping set pivots could be in principle solved by
considering the subgraph induced by the stopping set, and by
trying, for each variable node, whether the property expressed by
Definition \ref{def:pivot} is verified. However, there are some
complexity issues when following this approach. In fact, this
technique would be computationally onerous, especially for
stopping sets with large size, and when iteratively used as a
subroutine of some algorithm (like that one proposed in the next
section). The complexity of this pivot searching algorithm could
be reduced by exploiting the following lemma, which defines a
necessary condition for a variable node to be a stopping set
pivot.

\medskip
\begin{lemma}\label{lemma:necessary}
Necessary condition for a variable node belonging to a stopping
set $\mathcal{S}$ to be pivot for $\mathcal{S}$, is that the
variable node is connected to at least one check node with degree
2 in the subgraph induced by $\mathcal{S}$.
\end{lemma}
\begin{proof}
If no such check node is connected to the variable node, even if
the value of the variable node is known, no further correction can
be performed by the iterative decoder. In fact, after the
elimination from the graph of the variable node and of all the
edges connected to it, every check node continues to have a degree
at least 2.
\end{proof}
\medskip

According to Lemma \ref{lemma:necessary}, the search can be
restricted only to the variable nodes which are connected to at
least one check node of degree 2 in the subgraph induced by the
stopping set.

We propose next an alternative and more efficient algorithm for
the search of stopping set pivots. This algorithm is in general
not able to find all the pivots of a given stopping set, and its
success is bound to the condition that at least one pivot for the
stopping set is already available. However, for the purposes of
the optimization algorithm of LDPC on burst erasure channels
described in the next section, where two pivots for each stopping
set are always available, this algorithm comes out to be extremely
effective. The proposed pivot searching algorithm is based on the
following lemma. It defines a simple, sufficient condition for a
variable node to be a stopping set pivot.

\medskip
\begin{lemma}\label{lemma:sufficient}
Sufficient condition for a variable node $V_\alpha$ belonging to a
stopping set $\mathcal{S}$ to be pivot for $\mathcal{S}$, is that
there exists some $V_\beta \in \mathcal{S}$ and some check node
such that $V_\beta$ is pivot for $\mathcal{S}$, the check node has
degree 2 in the subgraph induced by $\mathcal{S}$ and it is
connected to $V_\alpha$ and $V_\beta$.
\end{lemma}
\begin{proof}
Let $C$ be the check node connected to $V_\alpha$ and $V_\beta$,
with degree 2 in the subgraph induced by $\mathcal{S}$. If the
value of the variable node $V_\alpha$ is known, and the value of
all the other variable nodes in $\mathcal{S}$ is unknown, then $C$
is capable to correct the variable node $V_\beta$. Since $V_\beta$
is a pivot of $\mathcal{S}$ by hypothesis, then all the variable
nodes of the stopping set will be corrected.
\end{proof}

\medskip
Combining Lemma \ref{lemma:necessary} and Lemma
\ref{lemma:sufficient} it is possible to prove the following
result.

\medskip
\begin{theorem}\label{theo:at-least-2-pivots}
If a stopping set $\mathcal{S}$ has pivots, then it has at least
two pivots.
\end{theorem}
\begin{proof}
Let the variable node $V_\alpha$ be a pivot of $\mathcal{S}$. For
Lemma \ref{lemma:necessary}, $V_\alpha$ must have at least one
connection towards a check node $C$, with degree 2 in the subgraph
induced by $\mathcal{S}$. Let $V_\beta$ be the second variable
node connected to $C$. From Lemma \ref{lemma:sufficient}, this is
sufficient to conclude that $V_\beta$ is a pivot of $\mathcal{S}$.
Thus, the number of pivots is at least two.
\end{proof}

\medskip
The variable nodes $V_\alpha$ and $V_\beta$ in the statement of
Lemma \ref{lemma:sufficient} will be referred to as neighboring
pivots. If it is known that the variable node $V$ is a pivot of a
certain stopping set, then all its neighboring pivots can be found
by looking, among the check nodes connected to $V$, for check
nodes with degree 2 in the subgraph induced by the stopping set.
Based on Lemma \ref{lemma:sufficient}, we propose the following
pivot searching algorithm for a stopping set $\mathcal{S}$ of an
LDPC code. As remarked above, the hypothesis is that at least one
pivot of the stopping set is already available at the beginning of
the algorithm.

\bigskip
\emph{Pivot Searching Algorithm.}
\begin{itemize}
\item $\,$[\emph{Initalization}] Set $\mathcal{P}^{(0)}$ equal to the available (non-empty) set of pivots
of $\mathcal{S}$. Set $\hat{\mathcal{P}}^{(0)} =
\mathcal{P}^{(0)}$.
%
\item $\,$[\emph{$\mathcal{P}^{(\ell)}$ expansion}] For each stopping set pivot
$V \in \hat{\mathcal{P}}^{(\ell)}$, apply Lemma
\ref{lemma:sufficient} in order to find the set
$\hat{\mathcal{P}}(V)$ of the neighboring pivots of $V$. Set
\begin{align}
\mathcal{P}^{(\ell+1)} = \Big( \bigcup_V \hat{\mathcal{P}}(V)
\Big) \cup \mathcal{P}^{(\ell)}.
\end{align}
Set \begin{align} \hat{\mathcal{P}}^{(\ell+1)} =
\mathcal{P}^{(\ell+1)} / \, \mathcal{P}^{(\ell)}.
\end{align}
\item $\,$[\emph{Stopping criterion}] If $\hat{\mathcal{P}}^{(\ell+1)}$
is equal to the empty set, stop and return $\mathcal{P}^{(\ell)}$.
Else, set $\ell = \ell + 1$ and goto the
\emph{$\mathcal{P}^{(\ell)}$ expansion} step.
\end{itemize}

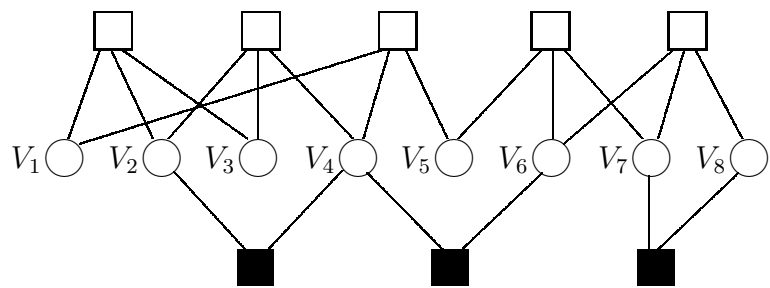
\begin{figure}[!h]
\centerline{
\unitlength .7mm 
\linethickness{0.4pt}
\ifx\plotpoint\undefined\newsavebox{\plotpoint}\fi 
\begin{picture}(138.65,54)(80,110)
\put(103.75,136.25){\circle{6.8}}
\put(85.25,136.25){\circle{6.8}}
\put(141,136.25){\circle{6.8}}
\put(122,136.25){\circle{6.8}}
\put(159.25,136.25){\circle{6.8}}
\put(196.75,136.25){\circle{6.8}}
\put(177.75,136.25){\circle{6.8}}
\put(215.25,136.25){\circle{6.8}}
\put(96.5,136){\makebox(0,0)[cc]{$V_2$}}
\put(78,136){\makebox(0,0)[cc]{$V_1$}}
\put(133.75,136){\makebox(0,0)[cc]{$V_4$}}
\put(114.75,136){\makebox(0,0)[cc]{$V_3$}}
\put(152,136){\makebox(0,0)[cc]{$V_5$}}
\put(189.5,136){\makebox(0,0)[cc]{$V_7$}}
\put(170.5,136){\makebox(0,0)[cc]{$V_6$}}
\put(208,136){\makebox(0,0)[cc]{$V_8$}}
\multiput(106,133.75)(.048076923,-.052447552){286}{\line(0,-1){.052447552}}
\multiput(142.75,133.5)(.048172757,-.049003322){301}{\line(0,-1){.049003322}}
\multiput(160.75,119)(.050664452,.048172757){301}{\line(1,0){.050664452}}
\put(196.25,133){\line(0,-1){14.25}}
\multiput(197.75,118.75)(.050488599,.048045603){307}{\line(1,0){.050488599}}
\put(155,112){\rule{7\unitlength}{7\unitlength}}
\put(194,112){\rule{7\unitlength}{7\unitlength}}
\put(118,112){\rule{7\unitlength}{7\unitlength}}
\multiput(138,134)(-.048076923,-.051282051){312}{\line(0,-1){.051282051}}
\put(91,157){\framebox(7,7)[]{}}
\put(119,157){\framebox(7,7)[]{}}
\put(174,157){\framebox(7,7)[]{}}
\put(145,157){\framebox(7,7)[]{}}
\put(200,157){\framebox(7,7)[]{}}
\multiput(86,140)(.048,.136){125}{\line(0,1){.136}}
\multiput(102,140)(-.04819277,.10240964){166}{\line(0,1){.10240964}}
\multiput(120,140)(-.067988669,.04815864){353}{\line(-1,0){.067988669}}
\multiput(105,140)(.048076923,.054487179){312}{\line(0,1){.054487179}}
\put(122,140){\line(0,1){17}}
\multiput(140,140)(-.048192771,.051204819){332}{\line(0,1){.051204819}}
\multiput(88,139)(.155080214,.048128342){374}{\line(1,0){.155080214}}
\multiput(142,140)(.04807692,.16346154){104}{\line(0,1){.16346154}}
\multiput(150,157)(.04819277,-.10240964){166}{\line(0,-1){.10240964}}
\multiput(160,140)(.048192771,.051204819){332}{\line(0,1){.051204819}}
\put(178,140){\line(0,1){17}}
\multiput(195,140)(-.048076923,.054487179){312}{\line(0,1){.054487179}}
\multiput(180,139)(.056149733,.048128342){374}{\line(1,0){.056149733}}
\multiput(198,140)(.04807692,.16346154){104}{\line(0,1){.16346154}}
\multiput(205,157)(.04812834,-.09090909){187}{\line(0,-1){.09090909}}
\end{picture}} \caption{Example
of subgraph induced by a size-8 stopping set with pivots
$\{V_2,V_4,V_6,V_7,V_8\}$. If $\mathcal{P}^{0} = \{V_2\}$, then
the set of pivots found by the pivot searching algorithm is
$\{V_2,V_4,V_6\}$.} \label{fig:pivot-searching-example}
\end{figure}

\bigskip
If only one pivot is available at the beginning of the algorithm
($|\mathcal{P}^{(0)}|=1$), since at least one neighboring pivot
must exist for the available pivot, the minimum number of pivots
returned by the algorithm is 2. At each step of the algorithm, the
sufficient condition expressed by Lemma \ref{lemma:sufficient} is
applied to the new pivots found at the previous step. The
algorithm is stopped as soon as no new pivots are found. For
instance, consider the stopping set of size 8 whose induced
subgraph is depicted in Fig. \ref{fig:pivot-searching-example},
where the variable nodes $V_2$, $V_4$, $V_6$, $V_7$ and $V_8$ are
supposed to be the stopping set pivots. In this figure, the check
nodes with degree 2 in the subgraph induced by the stopping set,
and connecting the neighboring pivots, have been depicted as
filled square nodes, while the other check nodes have been
depicted as non-filled square nodes. If $\mathcal{P}^{(0)} =
\hat{\mathcal{P}}^{(0)} = \{ V_2 \}$, then it follows
$\mathcal{P}^{(1)} = \{ V_2, V_4 \}$, $\hat{\mathcal{P}}^{(1)} =
\{ V_4 \}$, $\mathcal{P}^{(2)} = \{ V_2, V_4, V_6 \}$,
$\hat{\mathcal{P}}^{(2)} = \{ V_6 \}$, and $\mathcal{P}^{(3)} = \{
V_2, V_4, V_6 \}$, $\hat{\mathcal{P}}^{(3)} = \{ \, \}$. The set
of pivots $\mathcal{P}^{(2)}$ is returned by the algorithm. Note
that the pivots $V_7$ and $V_8$ cannot be found by the algorithm,
for $\mathcal{P}^{(0)} = \{ V_2 \}$.

This pivot searching algorithm is exploited in the optimization
algorithm for LDPC codes on burst erasure channels, presented in
the next section. The key for the application of the pivots
searching algorithm is the following observation: if for a given
LDPC code with maximum guaranteed resolvable burst length
$L_{\max}$, a burst of length $L_{\max}+1$ is non-resolvable, then
two pivots of the maximal stopping set included in the burst can
be always immediately found.

\section{Optimization Algorithm for LDPC Codes on Burst Erasure Channels}\label{sec:optimization-algorithm}
After having defined the concept of stopping set pivots, in this
section we present the LDPC codes optimization algorithm for burst
erasure channels.

For an LDPC code with maximum guaranteed resolvable erasure burst
length $L_{\max}$, any single erasure burst of length $L \leq
L_{\max}$ can be corrected by the iterative decoder independently
of the burst position within the codeword of length $n$. On the
contrary, there exists at least one erasure burst of length
$L_{\max}+1$, starting on some variable node $V_j$, which is
non-correctable by the iterative decoder. This implies that this
erasure burst includes some stopping sets. Next, it is proved that
the maximal stopping set included in the burst (defined as the
union of all the stopping sets included in the burst) has at least
two pivots. Specifically, the variable nodes $V_j$ and
$V_{j+L_{\max}}$, i.e. the first and the last variable nodes in
the burst, are pivots.

\medskip
\begin{theorem}\label{theo:first-last-pivots}
Let $L_{\max}$ be the maximum guaranteed resolvable burst length
of an LDPC code under iterative decoding, and let the erasure
burst of length $L_{\max}+1$ starting on the variable node $V_j$
and ending on the variable node $V_{j+L_{\max}}$, be
non-correctable. Then, $V_j$ and $V_{j+L_{\max}}$ are pivots for
the maximal stopping set included in the burst.
\end{theorem}
\begin{proof}
Let $\mathcal{S}$ be the maximal stopping set included in the
non-correctable erasure burst of length $L_{\max}+1$, and let
$\mathcal{G}$ be the subgraph induced by $\mathcal{S}$. By
hypothesis, an erasure burst of length $L_{\max}$, starting on the
variable node $V_j$ and ending on the variable node
$V_{j+L_{\max}-1}$, can be corrected by the iterative decoder.
This implies that if the value of the variable node
$V_{j+L_{\max}}$ is known, then the iterative decoder applied to
$\mathcal{G}$ is able to successfully correct all the variable
nodes in the maximal stopping set. Hence, $V_{j+L_{\max}}$ is a
pivot of $\mathcal{S}$.

Analogously, an erasure burst of length $L_{\max}$, starting on
the variable node $V_{j+1}$ and ending on the variable node
$V_{j+L_{\max}}$, can be corrected by the iterative decoder. By
reasoning in the same way, it is proved that $V_{j}$ is a pivot of
$\mathcal{S}$.
\end{proof}

\medskip
The theorem implies that the pivots' span of the maximal stopping
set included in the erasure burst of length $L_{\max}+1$ is equal
to $L_{\max}+1$. By combining Lemma \ref{lemma:non-disjoint} with
Theorem \ref{theo:first-last-pivots} we also obtain the following
result on the structure of the maximal stopping set included in a
non-correctable erasure burst of length $L_{\max}+1$.

\medskip
\begin{corollary}
Let $L_{\max}$ be the maximum guaranteed resolvable burst length
of an LDPC code under iterative decoding. Let the erasure burst of
length $L_{\max}+1$, starting on the variable node $V_j$, be
non-correctable, and let $\mathcal{S}$ be the maximal stopping set
included in the burst. Then, the subgraph $\mathcal{G}$ induced by
$\mathcal{S}$ is composed of non-disjoint bipartite graphs, i.e. a
path exists from any variable node in $\mathcal{G}$ to any other
variable node in $\mathcal{G}$.
\end{corollary}

\medskip
The optimization algorithm for LDPC codes on burst erasure
channels is described next. It receives an LDPC code parity-check
matrix in input, and returns an LDPC code parity-check matrix with
improved performance in environments where erasures occur in
bursts. This algorithm performs some permutations of the input
LDPC code variable nodes, in order to increase its maximum
guaranteed resolvable burst length, thus improving its burst
erasure correction capability. Neither the input code degree
distribution nor the connections between the variable and the
check nodes are modified by the algorithm. As a consequence, the
input LDPC code and the LDPC code returned by the algorithm have
the same degree distribution and the same performance on the
memory-less erasure channel.

\begin{figure}[!h]
\centerline{
\unitlength .7mm 
\linethickness{0.4pt}
\ifx\plotpoint\undefined\newsavebox{\plotpoint}\fi 
\begin{picture}(164,75)(80,100)
\put(84,126){\framebox(160,6)[]{}}
\put(124,126){\framebox(68,6)[]{}}
\put(127,129){\circle*{4.47}}
\put(189,129){\circle*{4.47}}
\put(144,129){\circle*{4.47}}
\put(161,129){\circle*{4.47}}
\put(214,129){\circle{4}}
\put(213,125){\vector(2,3){.1}}\put(190,125){\vector(-3,4){.1}}\qbezier(190,125)(202.5,109)(213,125)
\put(158,122){\makebox(0,0)[cc]
{$\underbrace{\phantom{ccccccccccccccccccccccccccccccc}}$}}
\put(158,114){\makebox(0,0)[cc]{$i$-th non-resolvable burst }}
\put(158,107){\makebox(0,0)[cc]{of length $L_{\max}+1$}}
\multiput(123.9,131.9)(0,.9583){25}{{\rule{.4pt}{.4pt}}}
\multiput(216.9,131.9)(0,.9583){25}{{\rule{.4pt}{.4pt}}}
\multiput(191.9,131.9)(0,.9583){25}{{\rule{.4pt}{.4pt}}}
\multiput(123.9,154.9)(0,.9375){17}{{\rule{.4pt}{.4pt}}}
\multiput(216.9,154.9)(0,.9375){17}{{\rule{.4pt}{.4pt}}}
\put(192,147){\vector(1,0){.1}}\put(124,147){\vector(-1,0){.1}}\put(124,147){\line(1,0){68}}
\put(174,175){\makebox(0,0)[cc]{pivots' span after}}
\put(157,156){\makebox(0,0)[cc]{pivots' span before}}
\put(174,169){\makebox(0,0)[cc]{swapping $> L_{\max} + 1$}}
\put(157,150){\makebox(0,0)[cc]{swapping $= L_{\max} + 1$}}
\put(217,165){\vector(1,0){.1}}\put(124,165){\vector(-1,0){.1}}\put(124,165){\line(1,0){93}}
\put(187,137){\makebox(0,0)[cc]{$V_p^{(i)}$}}
\put(212,137){\makebox(0,0)[cc]{$\tilde{V}^{(i)}$}}
\end{picture}} \caption{The pivot $V_p^{(i)}$ of
the stopping set $\mathcal{S}^{(i)}_{\max}$ is swapped with the
variable node $\tilde{V}^{(i)}$ (not in $\mathcal{B}^{(i)}$) in
order to make the pivots' span greater than $L_{\max}+1$.}
\label{fig:swap}
\end{figure}
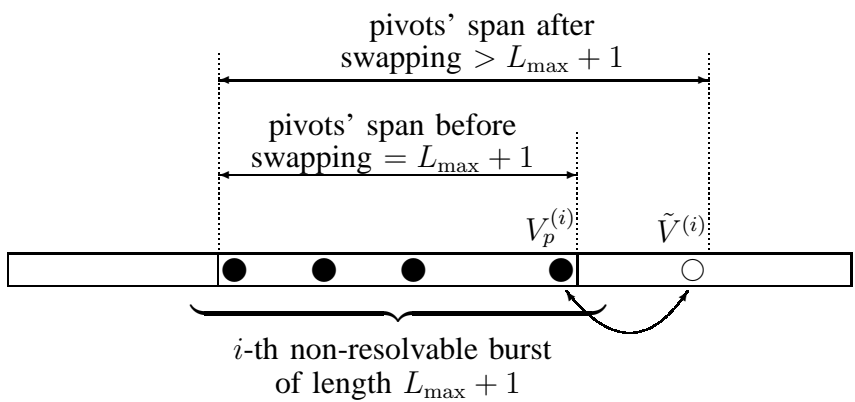

Consider an LDPC code with maximum guaranteed resolvable burst
length $L_{\max}$, let $\mathcal{V}$ denote the ensemble of all
its variable nodes, and suppose that the iterative decoder is not
able to successfully recover from a number $N_B$ of erasure bursts
of length $L_{\max}+1$ (some of the non-correctable erasure bursts
might be partly overlapped). Let $V^{(i,\,f)}$ and $V^{(i,\,l)}$
be, respectively, the first and the last variable node of the
$i$-th uncorrectable burst, with $i=1,\dots,N_B$, and let
$\mathcal{B}^{(i)}$ be the set of all variable nodes included in
the $i$-th burst. According to Theorem
\ref{theo:first-last-pivots}, the $i$-th uncorrectable burst
contains a maximal stopping set $\mathcal{S}^{(i)}_{\max}$ which
includes $V^{(i,\,f)}$ and $V^{(i,\,l)}$ among its pivots, and
other pivots of this stopping set can be eventually found by the
pivot searching algorithm presented in the previous section.
Suppose that one of these pivots is swapped with a variable node
$\tilde{V}^{(i)}$ not included in $\mathcal{B}^{(i)}$ such that,
after the swapping, the pivots' span of $\mathcal{S}^{(i)}_{\max}$
becomes larger than $L_{\max}+1$ (see Fig. \ref{fig:swap}). For
any starting erasure pattern given by an erasure burst of length
$L_{\max}+1$, the value of at least one pivot of
$\mathcal{S}^{(i)}_{\max}$ is now known, and the considered
stopping set will be resolvable for any possible position of such
burst.

If this procedure is applied to each uncorrectable erasure burst,
each maximal stopping set $\mathcal{S}^{(i)}_{\max}$
($i=1,\dots,N_B$) becomes resolvable for any possible burst
position. This does not necessarily imply that the erasure burst
length $L_{\max}+1$ will be resolvable at the end of the swapping
procedure, since any swap could in principle reduce the pivots'
span of some other stopping set. On the other hand, all our
numerical results reveal that this approach is indeed very
effective up to values of the erasure burst length $L$ extremely
close to $\lfloor p^* n \rfloor$.

If the sequence of $N_B$ variable node permutations makes the
erasure burst length $L_{\max}+1$ resolvable for any position of
the burst, then the new burst length $L_{\max}+2$ is considered.
On the contrary, a failure is declared, all the permutations are
refused, and a new sequence of $N_B$ permutations is performed.
The algorithm ends when a maximum number $F_{\max}$ of subsequent
failures is reached, for the same burst length. The $L_{\max}$
optimization algorithm is formalized in the following. The
algorithm input is an LDPC code with maximum guaranteed resolvable
burst length $L_{\max}$.

\bigskip
\emph{Pivot Searching and Swapping (PSS) Algorithm}
\begin{itemize}
\item$\,$[\emph{Initialization}] Set $L=L_{\max}+1$.
\item$\,$[\emph{Pivot searching step}] Set $F=0$. Find all the uncorrectable
erasure bursts of length $L$, and let the number of such bursts be
$N_B$. For each $i=1,\dots,N_B$, find all the pivots which are
neighbors of $V^{(i,\,f)}$ and all the pivots which are neighbors
of $V^{(i,\,l)}$. Let $\mathcal{P}_i$ be the set of pivots found
for the burst $i$.
\item$\,$[\emph{Pivot swapping step}] For each $i=1,\dots,N_B$:

1- Randomly choose a pivot $V_p^{(i)}$ in $\mathcal{P}^{(i)}$.

2- If $V_p^{(i)} \neq V^{(i,\,f)}$ and $V_p^{(i)} \neq
V^{(i,\,l)}$, then randomly choose a variable node
\begin{align}\label{eq:swapping}
\tilde{V}^{(i)} \in \mathcal{V}/ \Big( \mathcal{B}^{(i)} \cup
\big( \, \bigcup_{j \neq i}\mathcal{P}^{(j)}\, \big) \cup \{
\tilde{V}^{(1)},\dots,\tilde{V}^{(i-1)} \} \Big),
\end{align}
and swap $V_p^{(i)}$ and $\tilde{V}^{(i)}$.

$\phantom{--}$Else, if $V_p^{(i)} = V^{(i,\,f)}$, then randomly
choose a variable node $\tilde{V}^{(i)}$ as from
\eqref{eq:swapping}, and such that the index of $\tilde{V}^{(i)}$
is smaller than the index of $V^{(i,\,f)}$. Swap $V_p^{(i)}$ and
$\tilde{V}^{(i)}$.

$\phantom{--}$Else, if $V_p^{(i)} = V^{(i,\,l)}$, then randomly
choose a variable node $\tilde{V}^{(i)}$ as from
\eqref{eq:swapping}, and such that the index of $\tilde{V}^{(i)}$
is larger than the index of $V^{(i,\,l)}$. Swap $V_p^{(i)}$ and
$\tilde{V}^{(i)}$.

\item$\,$[\emph{Stopping criterion}] If the erasure burst length
is not resolvable, set $F=F+1$. If $F=F_{\max}$, stop and return
the new LDPC code with $L_{\max} = L-1$. If $F<F_{\max}$ goto
\emph{Pivot swapping step}.

$\phantom{--}$Else, if the erasure burst length $L$ is resolvable,
set $L=L+1$ and goto \emph{Pivot searching step}.
\end{itemize}

\bigskip
For each non-resolvable burst, the randomly chosen pivot
$V_p^{(i)}$ is swapped with a variable node $\tilde{V}^{(i)}$ in
order to guarantee that pivots' span of $\mathcal{S}^{(i)}_{\max}$
after the swapping is greater than $L_{\max}+1$. If $V_p^{(i)}
\neq V^{(i,\,f)}$ and $V_p^{(i)} \neq V^{(i,\,l)}$, then
$\tilde{V}^{(i)}$ is randomly chosen in $\mathcal{V} /
\mathcal{B}^{(i)}$, with the exclusion of the available pivots for
the maximal stopping sets of the other non-correctable bursts, and
of the variable nodes already swapped for the previously
considered bursts. If $V_p^{(i)} = V^{(i,\,f)}$, there are some
cases where the pivots' span of $\mathcal{S}^{(i)}_{\max}$ after
the swapping might be not greater than $L_{\max}+1$, i.e. when the
index of $\tilde{V}^{(i)}$ is larger than, and sufficiently close
to, the index of $V^{(i,\,l)}$. For this reason, if $V^{(i,\,f)}$
is selected, $\tilde{V}^{(i)}$ is chosen among the variable nodes
with index smaller than $V^{(i,\,f)}$. Analogously, if
$V^{(i,\,l)}$ is selected, $\tilde{V}^{(i)}$ is chosen among the
variable nodes with index larger than $V^{(i,\,l)}$.

The PSS algorithm is extremely flexible and can be in principle
applied to any LDPC code, independently of its structure, code
rate and codeword length. For instance, it can be applied to
either regular or irregular LDPC codes, both computer generated
(e.g. IRA \cite{jin00:IRA}, eIRA\footnote{There actually is a
difference between IRA codes and eIRA codes, albeit small. In
particular, the row weight for the systematic part of the
parity-check matrix is constant according to the definition of an
IRA code in \cite{jin00:IRA}. Further, according to
\cite{jin00:IRA}, the systematic part of the parity-check matrix
must be a (low-density) generator matrix for an IRA code (meaning
that it has more columns than rows). Neither of these contraints
are necessary for eIRA codes.}
\cite{ryan04:eIRA}\cite{ryan05:structured-eIRA} or GeIRA
\cite{liva05:CL} codes generated according to the PEG algorithm
\cite{xiao04:improved-peg,chen05:construction,hu05:peg},
protograph codes \cite{divsalar05:protograph,divsalar05:low-rate})
and algebraically generated (e.g. LDPC codes based on finite
geometries \cite{kou01:finite,lin01:class}). The optimized code
returned by the algorithm has the same performance as the input
code on the memory-less erasure channel, but is characterized by
an increased capability of correcting single erasure bursts. Then,
the PSS algorithm can be used within a two-step design approach,
consisting in first generating a good LDPC code for the
memory-less erasure channel, and then improving it for burst
correction. This approach leads to LDPC codes with good
performance in environments where the erasures are independent,
and where the erasures occur in bursts. The algorithm can be also
applied to already implemented LDPC codes: in this case, it can be
interpreted as a tool for the design of an \emph{ad hoc}
interleaver which will increase the robustness of the code to
erasure bursts.

\section{Numerical Results} \label{sec:numerical-results}
In this section, some numerical results on the $L_{\max}$
improvement achievable by applying the PSS algorithm are shown.
Five examples are provided. The first four examples are given for
LDPC codes with rate $R=1/2$ and different construction methods;
the fifth one for a rate $R=0.8752$ LDPC code. The four rate-$1/2$
codes are, respectively, a $(3,6)$-regular $(2640,1320)$ LDPC code
with Margulis construction
\cite{margulis82:explicit,mackay03:weakness}, an irregular
$(1008,504)$ LDPC code generated with the PEG algorithm, a
$(2000,1000)$ IRA code generated with the PEG algorithm and a
$(2048,1024)$ GeIRA code generated with the PEG algorithm. The IRA
and GeIRA codes construction was performed by first generating,
respectively, the double-diagonal and multi-diagonal part of the
parity-check matrix corresponding to the parity bits, and then
generating the systematic part with the PEG algorithm, also
considering the 1s already positioned in the parity part. The IRA
code is characterized by uniform check node distribution and by a
regular systematic part of the parity-check matrix, with all the
variable nodes corresponding to the systematic bits having degree
5. The GeIRA code is characterized by feedback polynomial (for the
recursive convolutional encoder) $g(D) = 1 + D + D^{420}$, and by
uniform check node distribution. The degree multiplicity for the
variable nodes corresponding to the parity bits is $1(1)$,
$419(2)$, $604(3)$, while the degree multiplicity of the variable
nodes corresponding to the systematic bits is $885(3)$, $85(13)$,
$54(14)$. Finally, the rate $0.8752$ code is a $(4,32)$-regular
$(4608,4033)$ LDPC code generated with the PEG algorithm (the
parity-check matrix for this code is $(256 \times 4608)$, with one
redundant row). The bipartite graphs of the $(2640,1320)$ Margulis
code and of the $(1008,504)$ irregular code are both available in
\cite{mackay:encyclopedia}, where the degree distribution of the
irregular code is also specified. The bipartite graph of the
$(2000,1000)$ IRA code, $(2048,1024)$ GeIRA code and high rate
code were generated independently.

The results obtained with the application of the PSS algorithm
(with $F_{\max} = n$) to the five codes are summarized in Table
\ref{tab:Lmax}. In this table, $n-k$ represents the maximum
possible value for $L_{\max}$, which can be obtained by generating
the parity-check matrix in burst correction form and applying to
it the quadratic complexity decoding algorithm, as explained in
\cite{fossorier05:universal}. For each code, the value of
$L_{\max}$ for the original code, and the estimate of the maximum
value of $L_{\max}$ achievable with column permutations ($\lfloor
p^* n \rfloor$, as suggested in \cite{ryan06:simplified}), are
also shown. At the bottom of the table, $L_{\max}^{\textrm{PSS}}$
denotes the maximum guaranteed resolvable burst length for the
code returned by the algorithm.

The starting value of $L_{\max}$ for the Margulis code (1033) was
already quite close to $\lfloor p^* n \rfloor$. On the contrary,
the values of $L_{\max}$ exhibited by the other codes (86, 403,
495 and 287 respectively) were quite poor with respect to $\lfloor
p^* n \rfloor$, especially for the irregular $(1008,504)$ PEG
code. When applied to the Margulis code, the PSS algorithm
returned a code with an excellent value of
$L_{\max}^{\textrm{PSS}}$, even larger than $\lfloor p^* n
\rfloor$. This result leads to a relevant conclusion: it is
possible to construct finite length LDPC codes with moderate
codeword length, such that $L_{\max} > \lfloor p^* n \rfloor$.
Furthermore, for the irregular PEG code, the IRA code, the GeIRA
code and the high rate regular code, the values of
$L_{\max}^{\textrm{PSS}}$ produced by algorithm were quite close
to (though lower than) $\lfloor p^* n \rfloor$. These examples
reveal the extreme effectiveness of the PSS algorithm for a wide
range of LDPC construction methods. As a comparison, in
\cite[Example 3]{tai06:burst}, a $(4,32)$-regular $(4608,4033)$
quasi cyclic LDPC code, whose construction is based on circulant
permutation matrices, is proposed for burst erasure correction.
This code is characterized by $L_{\max}=375$. As it can be
observed in Table \ref{tab:Lmax}, the original $(4,32)$-regular
code generated by the PEG algorithm has a value of $L_{\max}$
smaller than 375; however, the PSS algorithm was able to improve
this value beyond 375, up to 425.

It should be observed that for LDPC codes characterized by
systematic IRA-like encoding, the double-diagonal structure
allowing efficient encoding is lost due to the column permutations
executed by the algorithm. In this case, the variable node
permutation can be interpreted as an extra interleaving step to be
performed on the transmitter side prior of encoding. An
alternative approach for such codes consists in limiting the
algorithm permutations to the systematic variable nodes only, in
order to avoid the extra interleaving step. In this case, however,
the achievable values of $L_{\max}$ are smaller than those which
can be obtained by applying the permutation to all the
parity-check matrix columns. For example, we obtained $L_{\max} =
607$ for the $(2000,1000)$ IRA code of Table \ref{tab:Lmax}.

In order to give a more precise idea about the $L_{\max}$
improvement capability of the PSS algorithm, consider Table
\ref{tab:PPS-working}, where the details of how the algorithm
worked for the $(4,32)$-regular code are provided. In each table
entry two integer numbers are shown: the integer on the top is a
value of erasure burst length, while the integer on the bottom is
the number of uncorrectable positions registered for that erasure
burst length (denoted by $N_B$ in the formalization of the
algorithm proposed in the previous section). The burst lengths not
shown in the table are those ones for which $N_B=0$ was
registered. Hence, for instance, the first erasure burst length
that was recognized as non-resolvable for some burst position, was
$L=288$: for this length, one non-resolvable burst position was
found. By applying the pivot searching and swapping principle, a
column permutation was obtained which made the length $L=288$
resolvable for any burst position. Assuming this permuted version
of the parity-check matrix, the burst length $L=289$ was
investigated, and three burst positions where recognized as
non-resolvable ($N_B = 3$). Again, a column permutation was found
that guaranteed the length $L=289$ to be resolvable for any burst
position, and so on up to the burst length $L=426$, for which the
algorithm failed. As it can be observed from Table
\ref{tab:PPS-working}, for some values of $L$ the algorithm was
able to correct a relatively large number of uncorrectable burst
positions, e.g. $N_B = 12$ for $L = 395$ and $L=400$, $N_B = 13$
for $L = 389$ and $N_B = 18$ for $L = 403$.

\begin{table}[!t]
\caption{Original and improved values of $L_{\max}$ for the five
investigated LDPC codes.}\label{tab:Lmax}
\begin{center}
\begin{tabular}{||c|c|c|c|c|c||}
\hline \hline
 &   Margulis &   PEG irregular &   PEG IRA &   PEG GeIRA &   PEG regular\\
\hline \hline   $(n,k)$ &   $(2640,1320)$ & $(1008,504)$ &
$(2000,1000)$   &   $(2048,1024)$ &   $(4608,4033)$\\
\hline   $R$ &   $0.5$ &   $0.5$ &   $0.5$ &   $0.5$ &   $0.8752$ \\
\hline
  $n-k$ &   1320 &   504  &   1000 &   1024 &   575\\
\hline
  $L_{\max}$ &   1033 &   86  &   403 &   495 &   287\\
\hline
  $\lfloor p^* n \rfloor $ &   1133 &   473 &   861 &   946 &   445\\
\hline \hline
  $L_{\max}^{\textrm{PSS}}$ &   1135 &   446 &   852 &   914 &   425\\
\hline \hline
\end{tabular}
\end{center}
\end{table}

\begin{table}[!t]
\caption{Number of uncorrectable burst positions and corresponding
erasure burst lengths for the $(4608,4033)$ regular LDPC code
.}\label{tab:PPS-working}
\begin{center}
\begin{tabular}{||c|c|c|c|c|c|c|c|c|c|c||}
\hline \hline
288 & 289 & 290 & 300 & 322 & 328 & 330 & 334 & 335 & 337 & 340\\
1 & 3 & 3 & 1 & 1 & 1 & 1 & 1 & 1 & 1 & 1\\
\hline
344 & 352 & 354 & 356 & 361 & 362 & 365 & 368 & 373 & 374 & 376\\
1 & 1 & 1 & 1 & 2 & 1 & 1 & 1 & 3 & 2 & 8\\
\hline
377 & 378 & 379 & 380& 381 & 382 & 383 & 384 & 385 & 386 & 387\\
2 & 6 & 4 & 5 & 3 & 5 & 9 & 6 & 8 & 6 & 9\\
\hline
388 & 389 & 390 & 391& 392 & 393 & 394 & 395 & 396 & 397 & 398\\
5 & 13 & 6 & 2 & 5 & 9 & 7 & 12 & 3 & 9 & 7\\
\hline
399 & 400 & 401 & 402& 403 & 404 & 405 & 406 & 411 & 412 & 414\\
8 & 12 & 7 & 5 & 18 & 1 & 6 & 1 & 2 & 1 & 3\\
\hline
415 & 416 & 418 & 419& 420 & 421 & 422 & 423 & 424 & 425 & 426\\
3 & 2 & 2 & 2 & 3 & 5 & 4 & 6 & 5 & 4 & 9\\
\hline \hline
\end{tabular}
\end{center}
\end{table}
\section{Conclusion}\label{sec:conclude}

In this paper, a simple and effective algorithm for the
optimization of LDPC codes on burst erasure channels, under
iterative decoding, has been developed. The application of the
proposed algorithm to a given LDPC parity-check matrix leads to a
new parity-check matrix, characterized by properly permuted
columns, with a notable improvement in terms of maximum guaranteed
resolvable burst length. At each step of the algorithm, the
columns to be permuted are carefully chosen on the basis of a
local stopping set pivot analysis for the uncorrectable burst
positions. The optimized code has the same performance as the
original code on the BEC. Hence, if the input parity-check matrix
is optimized in order to achieve good performance on the
memory-less erasure channel, then the resulting code can be used
for communication both in scenarios with independent erasures and
in scenarios where the erasures occur in bursts, by only
exploiting the linear complexity iterative decoder. Numerical
results have been presented, showing the effectiveness of the
proposed approach for a wide range of LDPC code constructions.

\section*{Appendix\\ Algorithm Complexity}

In this appendix, we discuss some complexity issues for the PSS
algorithm. Specifically, given an LDPC code with maximum
guaranteed resolvable burst length $L_{\max} = L$, we evaluate the
complexity to obtain an improved code for which the erasure burst
length $L+1$ is resolvable.
As from the algorithm formalization given in Section
\ref{sec:optimization-algorithm}, we see that the operations
required for making the burst length $L+1$ resolvable are:
\begin{itemize}
\item[1.] Search of the unresolvable length-$(L+1)$ burst positions within the
codeword (their number is denoted by $N_B$).
\item[2.] Search of the pivots for each of the $N_B$ unresolvable
burst positions.
\item[3.] Selection of $N_B$ pivots.
\item[4.] Selection of $N_B$ variable nodes.
\item[5.] Variable node swapping.
\item[6.] Search of the possible unresolvable length-$(L+1)$ burst positions within the codeword.
\item[7.] Variable node swapping back if the erasure burst length
$L+1$ is non-resolvable.
\end{itemize}
The steps 1 and 2 are performed only once. Denoting by
$F_{\textrm{act}}$ the actual number of trials needed to obtain
the code for which the erasure burst length $L+1$ is resolvable,
the steps 3 to 6 are performed $F_{\textrm{act}}$ times, while the
step 7 is performed $F_{\textrm{act}}-1$ times (since the
permutation is accepted at the trial $F_{\textrm{act}}$).

The overall complexity is the summation of the complexities and is
dominated by the complexity of the steps 1 and 6, both requiring
to perform $n-L$ iterative decoding operations, each one with a
starting erasure pattern of size $L+1$. The complexity involved
with the other steps is negligible from a practical perspective,
due to the very simple operations performed (variable node
selection or variable node swapping) and to the fact that the
number $N_B$ of uncorrectable bursts of length $L_{\max}+1$ is
typically on the order of a few units, as shown for instance in
Table \ref{tab:PPS-working} for the $(4608, 4033)$ code. The
overall complexity is then given by the complexity required to
perform $(F_{\textrm{act}} + 1) (n-L)$ iterative decoding
operations, each one for $L+1$ unknown bits. The number
$F_{\textrm{act}}$ of trials needed to succeed is usually quite
small, typically ranging from a few units to a few tens. For
example, in the optimization of the $(2000,1000)$ code of Table
\ref{tab:Lmax}, $F_{\textrm{act}}$ was above 50 in one case only
and typically less than 10. Being the LDPC iterative decoder
complexity linear, the complexity required for generating a code
with $L_{\max} > L$ starting from a code with $L_{\max}=L$ is then
quadratic in the codeword length $n$.

For the codes presented in Table \ref{tab:Lmax}, the overall time
to obtain the code with $L_{\max} = L_{\max}^{\textrm{PSS}}$
starting from the original code ranged from less than three
minutes for the $(1008,504)$ code to about one hour and ten
minutes for the $(4608, 4033)$ code (whose parity-check matrix is
the least sparse).



\newpage


\vspace*{6 cm}

\vspace*{6 cm}

\vspace*{6 cm}

\vspace*{6 cm}
\end{document}